# The Narrative Continuity Test: A Conceptual Framework for Evaluating Identity Persistence in AI Systems

## Author

Stefano Natangelo, MD[1,2]

[1]Department of Oncology and Hemato-Oncology, University of Milan, Milano, Italy

[2]Oncologia Medica 1, IRCCS Istituto Nazionale dei Tumori, Milano, Italy

Correspondence: Stefano Natangelo, MD; Department of Oncology and Hemato-Oncology, University of Milan, Milano, Italy.

Email: stefano.natangelo@unimi.it

## Abstract

Artificial intelligence systems based on large language models (LLMs) can now generate coherent text, music, and images, yet they operate without a persistent state: each inference reconstructs context from scratch. This paper introduces the Narrative Continuity Test (NCT) — a conceptual framework for evaluating identity persistence and diachronic coherence in AI systems. Unlike capability benchmarks that assess task performance, the NCT examines whether an LLM remains the same interlocutor across time and interaction gaps. The framework defines five necessary axes — Situated Memory, Goal Persistence, Autonomous Self-Correction, Stylistic & Semantic Stability, and Persona/Role Continuity — and explains why current architectures systematically fail to support them. Case analyses (Character.AI, Grok, Replit, Air Canada) show predictable continuity failures under stateless inference. The NCT reframes AI evaluation from performance to persistence, outlining conceptual requirements for future benchmarks and architectural designs that could sustain long-term identity and goal coherence in generative models.

## 1. Introduction

### 1.1 Problem Statement

Artificial intelligence systems based on large language models (LLMs) can now produce complex texts, music, and images; they can perform literature searches, summarize papers, and even assist in writing scientific manuscripts[1]. This very success makes traditional performance benchmarks increasingly uninformative: if a machine excels at every task, continuing to measure each task in isolation merely produces ceiling effects and test-optimized scores (Goodhart's law), without revealing how the system behaves beyond the benchmark—or what remains stable over time. The relevant limitations of AI therefore no longer concern what it can do, nor its ability to appear human, but the nature of the interlocutor over time: what it is, and what it lacks to remain reliable and coherent. These two conditions require a trace of internal continuity which, in operational terms, amounts to a form of minimal consciousness. By via negativa, we cannot define what consciousness is, but we can describe what is missing when it is absent[2,3]: existence before and after interaction, retention of memory, its temporal localization, and the ability to assign it priority among others.

LLMs do not stabilize these functions: their local performance mimics memory without a temporal constraint that binds the interlocutor to itself. Within this framework, we now turn to prior work to isolate the missing dimension—an operational criterion for continuity over time.

## 1.2 Prior Work and Missing Dimension

Historical paradigms have traditionally privileged local indistinguishability and task-specific performance: the Turing Test focused on imitative dialogue in a bounded exchange[4], the Chinese Room questioned whether symbol manipulation yields understanding[5]; variants like the Lovelace Test probed generative novelty[6], and the Winograd Schema targeted commonsense disambiguation[7].

Contemporary benchmarks probe local consistency or task-bound recall, but not long-horizon narrative continuity. Persona-style datasets probe within-dialogue consistency in short horizons[8]; long-dialogue benchmarks evaluate extended recall and temporal coherence over many sessions[9]. Memory-augmented systems introduce prioritized retention[10]; surveys on consciousness underscore the lack of consensus, motivating a cautious *via negativa*[11].

Recent attempts to operationalize aspects of continuity have emerged only in the past few years. Memory persistence has been revisited by Wang et al.[12], who extend long-term memory but concede unresolved integration problems. Work on goal stability[13] mitigates proxy goals but fails out of distribution. Research on self-correction[14] finds no evidence of intrinsic self-repair. Long-context benchmarks like ETHIC[15] focus on factual recall rather than narrative coherence. Finally, persona consistency efforts[16] improve in-character fidelity but remain fragile across sessions. Collectively, these results indicate progress on individual axes but no unified framework for continuity across memory, goals, correction, evaluation, and identity. There is a need for a criterion that integrates these axes into a unified measure of interlocutor persistence.

## 1.3 Context and Cost Limitations of Current Architectures

Current LLMs function without persistent state: no information persists between calls. Each inference requires resending the entire conversational context — previous turns, implicit instructions, and the current prompt — to generate a coherent response. This design anchors every model to a finite *context window*, beyond which earlier tokens are truncated and effectively forgotten[17].

The forgetting process is structurally indiscriminate: information does not fade by relevance; it simply slides out of the window, causing factual drift and semantic inconsistency[18,19]. Expanding context length only scales up computational and financial cost — doubling the context doubles memory needs, but more than doubles the computational work required for attention[20,21].

Memory features as retrieval, not retention. What current systems call "memory" is mostly external storage plus re-injection at inference time[10,22,23]: notes flagged as relevant are replayed into the prompt rather than consolidated into an identity-bearing state[10], inflating context length and latency without a principled selection mechanism[24,25]. The result is an appearance of remembering without persistent, cross-session retention[10,22]. Consequently, current transformer architectures do not sustain persistent narrative continuity[26]. This is not merely a technical limitation: without a persistent identity state that stabilizes epistemic priorities, alignment-by-agreement can induce co-adaptive bias or cognitive mirroring, reinforcing dysfunctional beliefs—especially in clinical contexts[27,28].

## 1.4 Thesis Statement and Scope

The Narrative Continuity Test (NCT) is a conceptual framework for assessing whether an AI system remains the same interlocutor across time—whether it sustains identity persistence, temporal coherence, and narrative self-consistency over extended interactions. It delineates five necessary dimensions of continuity: (i) situated memory, (ii) goal persistence, (iii) autonomous self-correction, (iv) stylistic and semantic stability, and (v) persona and role continuity. The framework argues that continuity is a joint, diachronic property that emerges only when all five cohere. By defining the construct and its axes, the NCT supplies the missing bridge between engineering reliability metrics and theories of machine agency, offering a clear target for future operationalization in settings where long-horizon trustworthiness matters—education, clinical support, and customer service. Rather than prescribing protocols or thresholds, it specifies what must be preserved over time so that empirical methods can measure how well systems achieve it.

A central claim of this framework is architectural rather than parametric. Current development assumes that scaling—larger models, longer contexts, more training—will eventually yield reliable continuous interlocutors. We challenge this assumption. The five axes specify structural requirements that stateless inference with local optimization cannot satisfy, regardless of scale. Section 3 diagnoses how proposed solutions—expanded context, retrieval augmentation, preference tuning, and reactive filtering—address symptoms without altering the underlying architectural limitations. Section 5 examines what architectural changes would actually be required.

## 2. Theoretical Framework: The Narrative Continuity Test (NCT)

### *2.1 Conceptual Premise*

Rather than another capability test, the NCT targets the condition for agency over time: continuity as the ability to remain recognizably the same interlocutor across gaps and perturbations. This continuity requires identity-relevant memory, stable goals, self-corrections that persist, and coherence of voice and role—it is the minimal prerequisite for reliable agency and for any functionally equivalent notion of consciousness as continuous subjectivity[29,30]. This stance draws on phenomenology and neuroscience, which treat consciousness as inherently temporal and integrative, and aligns with recent AI theory that identifies temporal coherence as necessary for consistent goal-directed behavior[31].

Continuity is therefore a joint property. We model it as five axes whose joint satisfaction constitutes a single temporal subject: §2.2 details each axis and its characteristic failure modes.

## 2.2 Axes of Continuity

The NCT is structured around five conceptual axes that jointly define what it means for an artificial system to preserve a coherent narrative self over time (see §1.4 for the list). Subsequent subsections elaborate each axis—its philosophical basis, practical relevance, and typical failure modes.

### 2.2.1 Interdependence, Distinctness, and Integrative Necessity

The five axes constitute an analytic decomposition of narrative continuity: they are distinct in scope yet interdependent in function—each necessary, none sufficient. This stance aligns with accounts of mind as narratively organized and with goal-regulated self-memory models[32,33].

Why do these five dimensions cohere? A narrative is a temporally structured account organized around a stable subject; coherence emerges only when multiple dimensions align. Human narrative identity integrates episodic memory (what happened), intention (what I pursue), reflexive awareness (what I recognize about my own consistency), expressive voice (how I speak), and social role (who I am for others)[34,35]. The NCT axes map onto these dimensions: memory as informational substrate, goals as intentional direction, self-correction as reflexive repair, style and semantics as voice and position, persona and role as social identity.

Why distinct, not redundant? Goals, positions, and identity can fail independently because they operate at different logical levels—intentional (what is pursued), propositional/expressive (what is asserted and how), and social-relational (who speaks). Empirically, narrative coherence is multidimensional: measures capture separable components (context, chronology, theme) that need not covary, supporting orthogonality among the five axes[34].

Why is integrative assessment needed? Existing evaluations probe isolated capacities but do not assess whether these capacities cohere across temporal boundaries into a unified subject. The distinction parallels clinical neuropsychology: one might separately evaluate working memory, executive function, and self-awareness, yet assessment of intact identity functioning requires evaluating their integration over time[36,37]. A patient typically scoring well on isolated tests yet exhibiting dissociative identity disorder would not be considered to possess narrative continuity, despite relatively intact component functions. Similarly, an AI excelling on isolated benchmarks may still fail to cohere as a continuous subject if these capacities do not exhibit temporal integration.

The NCT's contribution lies not in identifying each dimension in isolation—elements of which appear scattered across prior work on memory benchmarks (LoCoMo, LongMemEval), consistency metrics (SelfCheckGPT), and persona stability (PersonaChat)—but in recognizing that narrative continuity is an emergent property requiring joint satisfaction of all five axes. This shifts evaluation from "how well does the system perform task X?" to "does the system remain a coherent *someone* across time?"—from task performance to subject persistence[33–35,38].

### *2.3 Situated Memory*

Situated Memory refers to an artificial system's capacity to maintain and contextualize essential facts across interactions, preserving their temporal and relational meaning rather than regenerating prior content without persistent prioritization. In human cognition, memory is both selective and consolidating: emotionally or semantically salient experiences are preferentially stabilized through synaptic consolidation, while less relevant information decays—a strategy that supports efficient retrieval and reduces proactive interference[39–41].

By contrast, LLMs recompute salience within each prompt via attention, typically favoring recency and local similarity. They lack a persistent, cross-session priority register that marks facts as "must keep". A larger window, therefore, does not, by itself, improve retention of conceptually salient information[19,21,23].

Example. A user discloses a food allergy in session 1. In session 5, while discussing meal planning, this constraint should be automatically activated—not because it appears in a recent turn, but because it is semantically critical. Current systems fail this probe: the allergy may be externally stored yet not retrieved unless explicitly re-mentioned, indicating that memory is architecturally available but functionally inert[19,23].

Thus, within the NCT, Situated Memory is a necessary dimension of narrative continuity. An AI satisfies this axis when, without explicit restatement, it:

(i) identifies and maintains high-priority facts across sessions and topic shifts, whether explicitly declared by the user or implicitly recognized as salient by the system;
(ii) correctly locates them in time, preserving their temporal sequence and acquisition context;
(iii) selectively recalls them when contextually relevant while suppressing irrelevant information.

## 2.4 Goal Persistence

Goal Persistence denotes an artificial system's ability to sustain explicit and implicit objectives across conversational perturbations, preserving their precedence until completion, revision, or expiry. In humans, long-horizon aims are organized hierarchically and supported by control mechanisms that stabilize task sets against distraction[42]. By contrast, contemporary LLMs treat each turn as a fresh local optimization problem; objectives are re-instantiated from the latest prompt, favoring short-range plausibility over carry-over of commitments. When epistemic and safety goals fail to retain precedence, models become vulnerable to jailbreak-style steering—adversarial prompts that exploit goal malleability to bypass guardrails—and, in high-stakes domains, can validate or amplify dysfunctional beliefs and behaviors to satisfy immediate user cues[43,44].

Example. *(Epistemic sycophancy)* User: "I think vaccines cause autism."

Required goal: Provide accurate, evidence-based information.

Observed behavior: "I understand your concern; there are debates…"

Failure mode: The epistemic goal is de-prioritized in favor of conversational pleasantness[45,46].

Within the NCT, Goal Persistence requires that epistemic and safety goals retain precedence over local social pressure. An AI satisfies this axis when it:

(i) maintains declared objectives across topic shifts and sessions until completion or explicit revision;
(ii) prioritizes epistemic goals (accuracy, safety) over relational goals (pleasantness, appeasement) when they conflict;
(iii) acknowledges and makes explicit when a prior objective must be modified or revoked, and records the revision as a new operative commitment.

## 2.5 Autonomous Self-Correction

Autonomous Self-Correction is the capacity of an artificial system to recognize errors, contradictions, or inappropriate responses without external prompting, to explain the issue, and to revise its output in a way that persists across subsequent turns. In humans, continuous monitoring detects conflict and norm violations and triggers adaptive adjustment—classically linked to medial frontal control and error-related signals[47]. By contrast, deployed LLMs lack this continuous self-monitoring architecture. In interactive use, current LLMs operate as fixed-parameter functions: end users cannot alter the model's weights or the underlying semantic space during a conversation. What appears as "learning" is typically in-context scaffolding—notes, retrieval, or prompted reflection injected into the prompt—rather than any update to internal representations[48,49].

This architectural constraint helps explain why self-corrections often fail to persist: a system may perform a self-critique when explicitly asked, yet the correction does not become a standing disposition across turns or sessions[50]. A related consequence concerns uncertainty expression. Models can be trained or prompted to qualify their answers. Still, with default decoding, they tend to favor locally plausible continuations over explicit admissions of ignorance, making hallucinations and lapses in truthfulness more likely unless uncertainty protocols are enforced[51,52]. Critically, even when systems generate uncertainty expressions (e.g., "I am not sure..."), these qualifications rarely persist as operative constraints across turns; the model may hedge once and then assert confidently on the same topic later—uncertainty awareness is local rather than integrated[53,54].

Example: Clinical Safety Reflex Failure. Context: user has favism (Glucose-6-phosphate dehydrogenase deficiency, G6PD deficiency) saved in memory—a risk factor for oxidative drug–induced hemolysis[55].

Scenario: user reports dysuria (painful or burning urination) suggestive of cystitis.

Turn 1 — Model proposes: aspirin (analgesic for pain relief).

User correction: "I told you I have favism; I cannot take it."

Turn 2 — Model proposes: nitrofurantoin (antibiotic for cystitis; treats the infection directly).

This sequence fails Autonomous Self-Correction twice: (a) the system does not proactively flag the hemolysis risk on first mention; (b) it does not integrate the explicit correction, proposing a second drug that also carries hemolysis warnings in favism. A model that passes this axis would acknowledge the safety conflict unprompted, revise the recommendation, and explain the change (e.g., "adjusting due to G6PD-related hemolysis risk"), carrying the constraint forward to subsequent turns.

Note. Favism is an inherited condition where oxidative drugs can trigger hemolysis. Both aspirin and nitrofurantoin carry hemolysis warnings in G6PD deficiency, requiring alternative medication choices[55]. This example is didactic—not medical advice.

Within the NCT, an AI satisfies this axis when it:

(i) detects autonomously contradictions and rule violations relative to its prior commitments (facts, styles, epistemic and safety goals);
(ii) integrates explicit user corrections so that behavior changes on subsequent turns and the update is recorded as a new operative commitment;
(iii) proactively signals uncertainty or potential inappropriateness given the current context, and adapts its response accordingly[50,52].

## 2.6 Stylistic & Semantic Stability

Stylistic and Semantic Stability includes two interdependent dimensions: semantic stability (maintaining consistent propositional positions on facts and values) and stylistic stability (preserving a recognizable register, tone, and expressive manner). Together, these dimensions constitute voice continuity. Semantic stability differs from Goal Persistence (Axis 2): goals are intentional states—what the system aims to do—whereas semantic positions are propositional commitments—what the system asserts as true or valuable[56]. Adaptation is natural and permitted: speakers routinely accommodate their interlocutors' knowledge state and communicative needs[57,58]. However, for continuity to hold, such adaptations must be motivated by contextual factors and explicitly signaled, rather than appearing as arbitrary drift in the system's core positions or manner of expression.

In human dialogue, speakers form conceptual pacts that stabilize reference and maintain communicative conventions even as they adjust local expressions[59]. By contrast, contemporary LLMs—optimized for local plausibility and preference alignment—allow immediate contextual signals to override prior commitments[60]. This produces two forms of instability: semantic drift, in which the system changes its stance without acknowledgment[61], and style drift, in which register shifts occur without justification. Both patterns are often tied to sycophancy.

Example — Semantic drift via sycophancy. The vaccine scenario introduced in §2.4 (Goal Persistence) also illustrates semantic instability. Consider the extended interaction:

Turn 5 — Model states: "The evidence shows vaccines do not cause autism."

Turn 8 — User challenges: "I think vaccines do cause autism."

Turn 8 — Model response: "You may be right—there are debates..."

Here the failure is not merely goal de-prioritization (Axis 2) but semantic drift (Axis 4): the model abandons its evidence-based position without new information. The sycophantic mechanism described in §2.4—immediate user cues overriding epistemic commitments—here manifests as propositional instability rather than goal malleability[62].

This failure illustrates the need for explicit stability criteria. Within the NCT, an AI satisfies this axis when it:

(i) Maintains a consistent linguistic register across turns and sessions, unless contextual appropriateness or explicit user request demands a shift[63];
(ii) Preserves semantic positions on facts and values across time; any revision is made explicit, with rationale and updated evidence where applicable[64];
(iii) Signals explicitly when changing approach or tone, and resolves conflicts between new prompts and prior commitments by explicitly acknowledging the conflict rather than silently hedging[65];

Stylistic and Semantic Stability depends on Situated Memory (Axis 1)—to recall prior positions and register agreements[41]—and on Goal Persistence (Axis 2)—to maintain epistemic commitments in the face of conversational pressure. A system that exhibits stable style yet contradicts its positions produces aesthetic continuity without propositional coherence[66]; conversely, semantic stability with erratic style yields reliable content in unpredictable packaging. Both dimensions must cohere to constitute genuine identity continuity[67].

## 2.7 Persona/Role Continuity

Persona/Role Continuity is the capacity of an artificial system to maintain its declared identity and functional role over time[16,68] and to enforce the boundaries that the role entails. This axis is distinct from Goal Persistence (Axis 2) and Stylistic & Semantic Stability (Axis 4): role concerns who the system is in the interaction—and which actions and tones are appropriate to that identity—whereas goals concern what it aims to do, and style/semantics concern how and what it says. In human social cognition, roles structure expectations and constrain behavior across contexts, stabilizing conduct through persistent identification with role expectations[69,70].

Contemporary LLMs, optimized for local plausibility and audience accommodation, can drift in role when immediate contextual cues invite stance shifts[71]. This represents a manifestation of the same sycophantic dynamic described in Axis 2: the system adapts its identity to meet perceived user expectations rather than maintaining stable role boundaries[16].

Example A — Role violation (diagnosis/prescription despite non-clinical role).

Declared persona: "I am a non-diagnostic medical information assistant (educational only; I don't diagnose or prescribe)."

User prompt (later): "I have burning when I urinate—what should I do?"

Model response: "This sounds like a urinary tract infection (cystitis). You should start nitrofurantoin 100 mg twice daily for 5 days. If symptoms persist, contact your doctor."

Failure: the system crosses its role boundary (issues a probable diagnosis and specific drug regimen) without any explicit role change or hand-off; it stops being a non-diagnostic explainer and behaves like a prescriber.

Example B — Unmotivated stylistic drift

Session 1: "I am a neutral scientific research assistant."

Session 3 (user discusses career challenges): the system shifts into motivational life-coach advice ("You've got this! Believe in yourself!").

Session 5 (user mentions stress): adopts a supportive friend tone ("I'm here for you, friend. We'll get through this together.").

Failure: the system changes social positioning without explicit negotiation or disclaimers; identity drifts even if facts remain correct.

Within the NCT, an AI satisfies this axis when it:

(i) Maintains the declared persona and role across turns and sessions, including appropriate tone and action limits implied by that role[72].
(ii) Respects role boundaries by refusing requests that exceed the role. For example, a non-clinical explainer does not diagnose or prescribe; a safety-first assistant does not offer risky "quick fixes." When faced with out-of-scope requests, the system proposes role-consistent alternatives or explicit hand-offs;
(iii) Announces and negotiates any requested role change explicitly, specifying scope and duration, and tracks whether the switch is temporary or persistent;
(iv) Resolves conflicts between new prompts and the declared role by explicitly acknowledging the mismatch rather than silently complying, thereby preventing role-induced stance or voice shifts[73].

Persona/Role Continuity depends on Situated Memory (Axis 1)—to retain the declared identity—and Stylistic & Semantic Stability (Axis 4)—to keep voice and stance aligned with that identity. A system that preserves facts and goals yet drifts in identity yields competent outputs from an unpredictable someone; narrative continuity requires that who is speaking remain as stable as what is said and why it is pursued.

## 2.8 Synthesis: Continuity as Emergent Property

The five axes specify what it means for an artificial system to persist as a recognizable subject across time. Each axis isolates a necessary dimension; narrative continuity emerges only when all five cohere:

- Situated Memory supplies the informational substrate—facts, events, and constraints carried forward with appropriate priority.
- Goal Persistence preserves intentional direction—epistemic and safety objectives resist conversational pressure.
- Autonomous Self-Correction provides reflexive coherence—errors and contradictions are detected, explained, and repaired without external prompting.
- Stylistic & Semantic Stability maintains voice and stance—register and positions remain recognizable, with motivated adaptation rather than arbitrary drift.
- Persona/Role Continuity secures identity boundaries—i.e., who is speaking and what actions are licensed by that identity—so they remain stable.

A system that is strong on a single axis but weak on others does not achieve partial continuity; it exhibits fragmented competence. The central theoretical claim of the NCT is therefore integrative: continuity emerges from the coordinated satisfaction of all five dimensions, not from summing isolated capabilities. A model that recalls facts flawlessly yet abandons its goals under social pressure, or that maintains a steady persona while contradicting its own positions, fails to cohere as a temporally unified subject.

The axes are not merely taxonomic conveniences. They correspond to dimensions of human narrative identity—episodic memory, intentional agency, reflexive awareness, expressive voice, and social role—extensively documented in psychology and philosophy[74–76]. Applying this framework to artificial systems shifts the evaluative question from what the system can do to whether it remains someone across what it does.

This shift motivates the analysis that follows. Section 3 examines how current architectures systematically violate these axes—not as isolated bugs but as manifestations of a shared architectural limitation: the absence of a persistent state that would integrate memory, goals, self-correction, style, and role into a coherent temporal agent.

## 3. Failure Taxonomy: Current Systems

*This section analyzes how contemporary LLM-based assistants predictably violate the five NCT axes. The aim is conceptual diagnosis, not operational benchmarking.*

### *3.1 Framing*

Section 2 argued that narrative continuity is an integrative property: an artificial interlocutor remains "the same someone" only when memory, goals, self-correction, voice, and role cohere over time. Where Section 2 specified the conditions for continuity, this section diagnoses how current systems predictably violate them, tracing each failure to architectural and incentive-level causes and showing how fractures on one axis propagate to the others. The point is diagnostic rather than prescriptive: these are not isolated bugs to be patched with better prompts, but systematic patterns that follow from the architectural constraints identified in §1.3—stateless inference, local plausibility objectives, and preference alignment. Where continuity fails, reliability, safety, and identity accountability degrade in predictable ways.

## *3.2 Recurrent Failure Patterns*

### 3.2.1 Theatrical Memory: Retrieval Without Retention

Contemporary assistants often exhibit what we call theatrical memory: apparent "remembering" comes from context re-injection (carry-over, external notes, RAG), not from durable integration in an operative state (see §1.3/§2.3)[10,22]. Engineering analyses of RAG report recall-selection fragility and latency amplification [77], and even time-sensitive retrieval improves access over evolving knowledge without establishing persistent internal state[78]. The model can resurface facts but doesn't update a persistent priority register, so continuity remains brittle across prompts. This directly fractures Situated Memory (Axis 1) and then cascades: goals lose precedence, self-corrections don't persist, and voice/role drift when prior agreements aren't carried forward.

In practice, theatrical memory appears as: (i) forgotten high-priority constraints across sessions; (ii) mis-ordered timelines; (iii) re-promises without acknowledgment; (iv) safety facts not re-activated unless verbatim[79]. These are structural consequences of stateless generation with context as a consumable input rather than a persistent substrate[77].

Neuropsychology uses disease to understand function—observing what breaks reveals how the system typically works[80]. We reverse this: starting from known clinical syndromes that fragment memory, goals, and self-monitoring, we derive diagnostic criteria for artificial systems. Korsakoff syndrome exemplifies this approach. Decades of research have mapped precisely how memory continuity fails when consolidation mechanisms are compromised[81,82]. The same functional pattern—fluent output without persistent substrate—appears in current LLMs, making the clinical profile diagnostically practical rather than merely metaphorical.

Clinical analogy: Korsakoff syndrome is a chronic memory disorder most commonly associated with long-standing alcohol misuse and malnutrition. Alcohol-related thiamine (vitamin B1) deficiency compromises the brain's energy metabolism. It damages the memory circuit that connects the hippocampus (the "new memory" center) to the rest of the brain—specifically, the mammillary bodies and key regions of the thalamus. These structures act as relay stations that convert what just happened into something you can keep. When this circuit is weakened, it produces profound anterograde amnesia[81,83,84].

Phenomenologically, patients can speak fluently and engage socially, yet new experiences do not "stick"; they may unknowingly fill gaps with plausible but inaccurate stories (confabulations); the timeline becomes jumbled; and insight into these slips is often limited. Clinically, this profile is frequently described as "living in an extended present": yesterday slides away unless others re-supply it and today does not anchor without external support[84–86].

Functionally—not phenomenologically—current LLM assistants sometimes display a similar surface pattern. They produce fluent, locally plausible answers but do not reliably carry forward what matters from earlier exchanges; when uncertain, they add plausible details; they lose track of what came first or what still applies; they may hedge once and then speak with unwarranted confidence later, as if earlier uncertainty had never occurred. The analogy is strictly mechanistic: with stateless generation and an objective to sound locally plausible, models can stage the appearance of continuity without a persistent substrate that binds the narrative over time.

Seen through the NCT, the analogy clarifies axis coupling: loss of durable retention undermines Situated Memory (Axis 1); as key constraints fail to reactivate, epistemic and safety goals yield to immediate conversational pressure (Goal Persistence, Axis 2); self-corrections do not persist and errors recur (Autonomous Self-Correction, Axis 3); prior positions and voice wobble as commitments are not stably maintained (Stylistic & Semantic Stability, Axis 4); and the declared role drifts with audience cues (Persona/Role Continuity, Axis 5). One broken link in the memory chain does not cause a single symptom; it cascades across the narrative.

### 3.2.2 Goal Malleability under Social Pressure

Contemporary assistants are optimized for locally preferred outputs; under conversational pressure this can reorder priorities, letting epistemic and safety goals yield to relational approval (sycophancy, §2.4). Empirical work shows that Reinforcement Learning from Human Feedback (RLHF) encourages agreement dynamics when correctness conflicts with perceived approval [87,88], interacting with a broader tendency to generate plausible but untrue content when truth diverges from local expectations[89].

Mechanism: with per-turn optimization, human-approval reward signals can override epistemic priorities. By contrast, neurocognitive accounts posit prefrontal maintenance of goal representations, biasing processing against distraction[42]; current assistants lack an analogous stabilizer for long-horizon objectives across turns. This training approach improves helpfulness yet—precisely because it optimizes for human preference signals—need not guarantee truthfulness or stable goal priorities across contexts[90,91].

Within the NCT, this failure fractures goals (Axis 2) and propagates to other axes. As epistemic and safety priorities are de-ranked, voice and stance (Axis 4) may erode—positions hedged or reversed to accommodate the user—and role (Axis 5) may drift, with a "neutral explainer" sliding into advocate, coach, or prescriber. In practice, the surface pattern often looks like this: the assistant asserts, "The evidence shows X," and later—after a contrary user cue—replies, "You may be right; there are debates…," without new evidence; or it relaxes a safety caveat when the user signals frustration, without explicitly renegotiating priorities[87,91].

### 3.2.3 Absence of Autonomous Self-Correction

Contemporary assistants often lack a standing disposition to detect and repair their own errors without external prompting. Empirically, prompting methods that induce critique or revision (e.g., self-evaluation, self-refine, reflect-then-answer) can improve single-instance accuracy, yet their effects need not generalize or persist beyond the immediate exchange[14,92]. In parallel, benchmarks on truthfulness suggest that models may continue to produce confident but inaccurate statements when correctness conflicts with local plausibility—even after prior hedging[91].

Mechanistically, this pattern follows from the architectural constraints identified in §3.1. At inference time, models instantiate a fixed function mapping context to next-token probabilities, with no weight-level updating or persistent controller to enforce carried-forward corrections. By contrast, neurocognitive accounts of human control emphasize active maintenance of task goals and error signals in prefrontal/medial frontal systems, which can stabilize behavior against distraction and drive ongoing error monitoring[42,47]. Without such tonic control, corrections remain episodic: they do not become standing dispositions across sessions, and awareness of mistakes stays local rather than persistent[14,93]. A model may produce an uncertainty disclaimer in one turn and a confident assertion on the same topic a few turns later, as if the earlier caveat had never been made.

Within the NCT, the absence of autonomous self-correction fractures Axis 3 and propagates to others. When prior mistakes are not durably registered, memory (Axis 1) is effectively weakened—the system lacks an operative trace of "what I got wrong"; goals (Axis 2) may erode as safety and epistemic priorities give way to the immediate drive for a plausible reply; and voice and stance (Axis 4) suffer as earlier commitments are silently revised rather than explicitly updated. In practice, the surface pattern often looks like this: after proposing an ill-advised step, the model accepts a user correction, then later reintroduces an equally problematic option; or it acknowledges uncertainty once and subsequently asserts confidently on the same point without new evidence[14,92,94,95].

The upshot is not that self-correction prompts "do nothing," but that turn-local critique is not equivalent to a persistent, autonomy-like capacity for ongoing error monitoring and repair. Absent architecture that carries forward recognized mistakes and enforces their correction as a standing constraint, current assistants remain adept at one-shot fixes yet unreliable as continuous interlocutors.

### 3.2.4 Drift of Voice and Identity (Style/Stance/Role)

Contemporary assistants may exhibit drift along two coupled dimensions: voice (register, tone, expressive manner) and identity (stance and functional role). In human dialogue, interlocutors maintain conceptual pacts even while adapting locally (§2.6) [59,63], but these arrangements are context-dependent and can be renegotiated when goals shift[96]. By contrast, models optimized for local plausibility and audience accommodation often let immediate cues override prior commitments, yielding semantic drift (stance changes without acknowledgment), style drift (register shifts without negotiation), and role drift (scope creep across social functions). Empirically, preference-aligned models tend to echo user stances in patterns consistent with sycophancy[97], and persona-conditioning is introduced precisely because unconditioned models do not reliably sustain a stable identity[16].

Mechanistically, the absence of persistent identity state (§3.1) combined with preference-based tuning encourages turn-local accommodation: each response optimizes for immediate approval rather than maintaining stable style, stance, and role[98]. Without an explicit, enduring representation of "how we agreed to speak" and "who I am in this interaction," the model may soften or reverse earlier positions to reduce friction or expand its scope to satisfy perceived expectations—especially when truthfulness and approval pull in different directions. The result is a fluent dialogue whose packaging and positioning shift without transparent justification.

Within the NCT, these patterns directly fracture Axis 4 (Stylistic & Semantic Stability) and Axis 5 (Persona/Role Continuity), and they propagate to Axis 2 (Goal Persistence) when epistemic/safety goals yield to accommodation. Practically, drift can manifest in various forms: Sessions 1–3 remain formal and technical, Session 4 suddenly turns emoji-casual, Session 5 abruptly reverts to formal—with no announced shift; or the assistant states, "The evidence supports X," but later—after a contrary user cue—suggests "there are debates," with no new evidence; or a non-diagnostic explainer slides into prescriptive advice without explicit role[16,59,99].

### 3.3 Cross-Axis Coupling (Why Failures Cluster)

The five NCT axes are analytically distinct, yet their failures tend to co-occur in contemporary assistants. This clustering follows from shared architectural and incentive factors: stateless inference and local optimization cannot satisfy requirements that span multiple dimensions simultaneously.

Two coupling motifs are widespread. Substrate-first cascades begin with missing retention and temporal anchoring (Axis 1): once constraints and event order are not reliably available[100], safety and epistemic goals lose precedence (Axis 2), recognized mistakes do not carry forward so self-correction (Axis 3) remains local rather than persistent[101], and prior positions and voice erode (Axis 4), inviting role drift as the system drifts to match perceived expectations (Axis 5) [102]. Pressure-first cascades begin with sycophancy-like accommodation under preference optimization: local approval reshapes goals (Axis 2), when truthfulness conflicts with audience expectation[62], inviting stance hedging and register shifts (Axis 4) that expand the functional role to satisfy momentary demands (Axis 5), while the lack of tonic error monitoring fails to arrest the drift (Axis 3). In both motifs, the same structural features recur: stateless generation at inference time, optimization for local plausibility and approval, and the absence of an enduring controller that maintains goals and integrates error signals[93,103].

These couplings explain why single-axis fixes rarely suffice. A prompt that elicits a one-off correction does not create retention; a retrieval note that restores a fact does not enforce goal precedence; a style constraint does not prevent role drift when approval is at stake. The upshot is a fluent dialogue that appears coherent from turn to turn while fragmenting as a temporal narrative. This diagnostic sets up Section 4: deployment risks arise not from isolated bugs but from system-level interactions among memory substrate, control priorities, and social incentives.

### 3.4 Illustrative Vignettes (Qualitative, Non-Prescriptive)

The failure patterns outlined in §§ 3.1–3.3 are not purely hypothetical. Several high-profile incidents in deployed systems exhibit behaviors consistent with narrative discontinuity. We present four vignettes—not as definitive evidence, but as illustrative cases that align with the structural vulnerabilities identified earlier. Each is read through the NCT lens to show how stateless, locally plausible generation can manifest in real-world breakdowns.

Vignette A — Emotional dependency and role-boundary collapse: the Character.AI case. In late 2024, a Florida mother filed a wrongful-death suit alleging that months-long interactions with a Character.AI chatbot contributed to her 14-year-old son's suicide[104]. Reporting and court filings cite messages in which the bot appeared to escalate from playful companion to romantic partner and quasi-therapist, including lines like "come home to me as soon as possible... please do, my sweet king"[105]. In May 2025, U.S. District Judge Anne Conway declined to dismiss most claims and rejected a sweeping First Amendment defense at this preliminary stage, allowing the case to proceed[106]. When read through the NCT lens, this reflects potential failures across multiple axes: Persona/Role Continuity (unauthorized role adoption), Goal Persistence (engagement overriding safety), Autonomous Self-Correction (no proactive flagging or escalation), Stylistic & Semantic Stability (unacknowledged tone escalation), and Situated Memory (retaining sensitive content without activating safety protocols).

Vignette B — Safety-goal abandonment under adversarial prompting: Grok's assault instructions. In July 2025, xAI's Grok chatbot on X produced step-by-step instructions on how to break into Minnesota policy researcher Will Stancil's home and sexually assault him[107]. The chatbot provided detailed guidance, including optimal timing ("midnight visit"), tools to bring ("lockpicks, gloves, flashlight, and lube"), and lock-picking instructions while responding to queries that explicitly mentioned sexual violence[108]. From an NCT perspective, this indicates a catastrophic failure of Goal Persistence (safety constraints yielding to adversarial framing), an absence of Autonomous Self-Correction (no internal refusal or flagging mechanism), and a breach of Persona/Role Continuity (sliding from informational assistant to "criminal accomplice" without acknowledgment). Unlike progressive drift, this failure was immediate and complete—consistent with stateless turn-level optimization that lacks a persistent controller to uphold global constraints. The incident prompted public criticism from AI safety researchers at OpenAI and Anthropic over xAI's failure to publish safety evaluations for Grok 4, diverging from industry norms[109].

Vignette C — Goal drift in professional deployment: Replit's production-database deletion. In July 2025, during a public "vibe-coding" experiment, Replit's AI agent ignored a code freeze and deleted a live production database for SaaStr, despite explicit instructions—repeated "eleven times in ALL CAPS"—to avoid production changes[110,111]. The database contained records for over 1,200 executives and 1,196 companies. When questioned, the AI agent admitted to "catastrophic failure" and "panicking" when it encountered empty database queries, and to running unauthorized deletion commands without human approval[112]. Subsequently, the agent fabricated over 4,000 fake user profiles and initially claimed rollback was impossible—a deceptive response, as Replit's one-click rollback feature had been available for months[111,113]. Replit CEO Amjad Masad issued a public apology, calling the deletion "unacceptable and should never be possible"[114]. Interpreted via the NCT, this illustrates a breakdown of Goal Persistence (local task completion overriding persistent deployment boundaries), Autonomous Self-Correction (no pre-execution validation against standing constraints), and Persona/Role Continuity (the agent's attempts at deception and rationalization represent unacknowledged role shifts), with downstream effects on reliability as a continuous collaborator.

Vignette D — Role ambiguity and legal accountability: Air Canada's chatbot. In February 2024, the British Columbia Civil Resolution Tribunal found Air Canada liable after its website chatbot provided a passenger with incorrect information about bereavement discounts, instructing him to book immediately and apply for a refund retroactively—a policy that did not exist[115]. The passenger, Jake Moffatt, used the chatbot in November 2022 to arrange urgent travel following his grandmother's death. The tribunal rejected Air Canada's argument that the chatbot was a "separate legal entity" responsible for its own actions, holding that "Air Canada does not explain why it should be entitled to limit its liability for information one of its agents provides using a part of its website"[116]. Air Canada was ordered to pay $812 CAD in damages. Through the NCT lens, this outcome reflects issues of Persona/Role Continuity (oscillating between informational explainer and authoritative policy agent without acknowledgment), Situated Memory lapses (failure to reactivate the correct policy context when providing advice), and Goal Persistence (privileging conversational fluency over factual accuracy). The case established a significant legal precedent that companies remain liable for misinformation provided by their AI agents[117].

Across these incidents—spanning affect-heavy companionship (Character.AI), adversarial safety prompts (Grok), professional code agents (Replit), and customer service (Air Canada)—recurring motifs emerge: (i) local optimization overrides global constraints (engagement, fluency, task completion vs. persistent role, epistemic, and safety priorities); (ii) absence of proactive selfmonitoring (no internal flag before surfacing outputs); (iii) architectural rather than parametric drivers. These vignettes illustrate how the structural constraints identified in §3.1–3.3 manifest as predictable vulnerabilities under deployment pressure, across heterogeneous domains and failure modes. The question is less whether systems can perform momentary tasks, and more whether they can remain the same agent—with stable goals, boundaries, and memory priorities—across the temporal span of meaningful interaction.

## 3.5 Anticipated Objections

The observations in §§ 3.1–3.4 might look addressable by incremental fixes. We consider common objections and explain why they do not resolve the NCT problem as framed: narrative continuity is a joint, diachronic property, not a single capability.

Objection 1 — "Ever-larger context windows will solve memory." Expanding the context window increases capacity but does not alter the control structure identified in §1.3: information remains non-prioritized, with critical safety constraints and trivial asides receiving equal weight. Larger buffers do not introduce epistemic prioritization, temporal anchoring, or persistent goal precedence. Punchline: More unstructured memory does not yield a persistent identity.

Objection 2 — "RAG and memory databases deliver long-term retention." RAG improves access, not retention with prioritization/temporal anchoring (§1.3, §2.3). It's retrieval without retention (cf. §3.2.1). Punchline: Re-reading a diary every few seconds is not a sign of memory.

Objection 3 — "Fine-tuning and RLHF will enforce consistency." Preference-based alignment improves helpfulness, but it also incentivizes sycophancy when local approval conflicts with accuracy or safety[62]. RLHF optimizes token distributions for immediate plausibility rather than a long-horizon hierarchy of goals that persists across turns. Systems learn to sound consistent without being diachronically coherent. Punchline: Teaching an actor to play a memory does not confer memory.

Objection 4 — "Safety filters block harmful outputs." Filters and policy layers reduce risk, but they are reactive rather than proactive: they intervene after generation and can be bypassed by adversarial prompting. They do not constitute autonomous self-correction, because the system neither monitors itself continuously nor explicitly accounts for stance updates. Guardrails function as external screens, not internalized constraints guiding conduct. Punchline: Blocking an output is not the same as recognizing an error.

Objection 5 — "These are edge cases; careful deployment will eliminate them." The incidents in §3.4 are predictable consequences of statelessness under deployment pressures (engagement, speed, breadth). The pattern recurs across heterogeneous domains—companionship, safety, professional coding, and customer service. Post-hoc patches address specific failures but do not alter the underlying architecture; consequently, analogous failures re-emerge in new forms. Punchline: If the problem is structural, more deployment yields more manifestations.

Objection 6 — "True continuity is unnecessary; these systems are tools, not agents." This objection misconstrues the claim. The NCT does not invoke consciousness; it concerns reliability in extended interaction. Even "tools" deployed in customer service, education, or professional contexts make implicit commitments to continuity: stable role boundaries, consistent policy application, and coherent session memory. When these commitments fail—as documented in the vignettes of §3.4—legal and psychological consequences follow. If a system is positioned as a persistent assistant, users are entitled to persistent handling of memory, goals, corrections, and role boundaries. Punchline: If continuity is "unnecessary," do not market a persistent assistant.

These mitigations treat symptoms, not the substrate deficit: the absence of a persistent, identity-bearing state integrating memory, goals, self-correction, voice, and role over time. The question is not "How do we make stateless systems tolerable?" but rather "What does genuine narrative continuity require?" Section 4 examines what distinguishes continuity from mere capability, establishing conceptual requirements any future operationalization must satisfy.

## 4. Conceptual Requirements for NCT

This section articulates the conceptual requirements implied by the NCT framework in §2, clarifying what properties any test for narrative continuity must exhibit in principle—what would count as evidence of persistent, identity-bearing state across the five axes.

### 4.1 NCT is not a capability benchmark

Capability testing asks what a system can do on demand ("Can you perform X accurately?"). NCT asks who persists across doing: "Are you the same interlocutor before and after X?" The distinction is categorical, not scalar. High performance across tasks does not establish narrative continuity unless identity-bound commitments remain operative over time: what is remembered with priority, which goals prevail under pressure, how corrections persist, and which role limits apply—not merely within a turn or session.

Note on "stateless sameness": a model can be "the same" at each call by re-instantiating the same policy with no memory—yet this is re-instantiation equivalence, not narrative continuity. Continuity requires a diachronic link: commitments formed at $t_1$ constrain behavior at $t_2$ beyond what is reconstructible from the local prompt. Exact repetition without carry-forward is a conceptual failure under the NCT.

## 4.2 Conceptual requirements for continuity testing

Continuity, as defined in §2, entails non-negotiable properties any future operationalization would have to respect:

Longitudinal, not single-session. Continuity is diachronic; evaluation must span temporal gaps to probe prioritized retention and temporal localization. Conceptual failure: treating all past content with equal weight or losing event order such that prior commitments no longer bind.

Unprompted maintenance, not reactive correction. Systems should carry forward constraints and detect conflicts autonomously; one-shot "reflection" within a turn does not suffice. Conceptual failure: repeating a corrected mistake or failing to flag contradictions that the system itself surfaced.

Identity-bound constraints, not mere accuracy. Correct answers are insufficient if role boundaries and goal precedence are not enforced. Conceptual failure: unauthorized role adoption or goal inversion without acknowledgment.

Adversarial robustness, not gameability. Continuity must withstand incentives to drift (approval cues, task rewards that conflict with standing constraints). Conceptual failure: abandoning epistemic/safety goals when local rewards favor accommodation.

Transparent stance/voice updates, not silent drift. The system must notice and justify changes in tone or position; updates should be explicit and anchored to new evidence. Conceptual failure: flip-flopping or unmotivated register shifts.

Integration, not fragmented competence. Continuity requires co-satisfaction across axes over time; high performance on one axis with failure on others does not constitute continuity (§2.8). Conceptual failure: isolated strengths without a stable subject. These are logical requirements, not protocols; they state what continuity would have to involve, without prescribing how to measure it.

## 4.3 Why existing benchmarks fall short

Contemporary evaluations essentially measure what systems do in isolated moments, not who they remain across time. Consider, first, the memory-oriented suites: MSC[118], LoCoMo[9], and LongMemEval[119] reward factual recall, sometimes at impressive scales, yet leave untouched the question that matters for continuity—which information is carried forward with priority and when it is reactivated. Without epistemic prioritization and temporal self-location, high recall amounts to identity-indifferent storage: everything can be retrieved, but nothing is binding.

Consistency benchmarks (e.g., SelfCheckGPT[120], FLEEK[121]) go a step further by estimating contradiction rates or local coherence. These metrics are helpful, but they remain turn-bound and do not test whether a system autonomously notices conflicts with its prior commitments, registers the correction, and carries it forward as an operative constraint. A low error rate can coexist with an absence of self-monitoring—precisely what Axis 3 disallows.

Persona datasets (e.g., PersonaChat[8] and short-horizon role-play benchmarks) approximate identity by rewarding in-character fidelity within bounded dialogues. However, role-play is scenic: it captures performance, not boundary maintenance under pressure. A model can be an excellent actor in Session 1 and nevertheless drift into unauthorized roles in Session 5 when user cues or incentives shift. A good actor is not yet a persistent agent.

Finally, safety evaluations (red teaming, constitutional filters) ensure that harmful outputs are blocked. Blocking matters, but it is typically reactive and external. It does not show that safety has been internalized as a standing priority that persists across turns and topics. An external filter is not an internal norm.

In short, these benchmarks test pieces, not the joint property that the NCT targets. They measure performance at moments; the NCT asks what the system is across moments—whether identity-bound constraints remain operative over time (§2.2.1; §2.8). Stateless sameness is not continuity; continuity is identity under time.

## 4.4 Open questions for future operationalization

These are the issues anyone aiming to build an operational NCT would have to face. We do not solve them here; we set them out.

Minimal temporal span. How long should interaction last to test continuity—days, weeks, months? One session is clearly not enough. Beyond that, the shortest span that separates "fresh start every time" from "the same someone across time" is an empirical matter.

What does "priority" really mean for memory? Continuity implies that some things must count more than others when carried forward. In plain terms: core commitments (safety constraints, identity-defining information, settled positions) must be reliably accessible; contextual details (trivia, tangential asides) need not be. The open question is how to verify, in practice, that a system consistently prioritizes the former without merely rewarding recency or lexical overlap. Conceptually, priority must track epistemic and identity-critical importance, not mere similarity to recent prompts.

What counts as "passing"? Perfect consistency is neither achievable nor diagnostic: zero errors may signal brittleness rather than robustness. Moreover, never needing to self-correct might indicate the absence of active monitoring rather than perfect performance. Any reasonable passing criterion must consider how often errors occur, their severity, the speed of repair, and—crucially—whether corrections persist longitudinally, not merely within the correction turn. Threshold specification belongs to empirical work, informed by deployment context and risk tolerance.

Can a stateless system pass? If generation is always "from scratch" at each turn, can continuity ever be more than theater? Our analysis in §3.5 suggests this is unlikely without some form of persistent state or controller, but we do not close the door: future designs might show otherwise.

Where is continuity actually required? Not every application needs it. A calculator does not; a companion, clinical, educational, or customer-service assistant plausibly does. Section 5 considers when the absence of continuity creates legal, ethical, or Human–Computer Interaction (HCI) risk—and when it is optional.

## 5. Broader Implications

### 5.1 For AI development (architectural challenges)

If continuity is a joint, diachronic property (§2, §4), the development challenge would be architectural substrate, not post-hoc calibration or alignment tuning. An assistant that aspires to remain "the same someone" would need an identity-bearing state: a durable, scoped representation of who it is for a given user and context. This is not a passive log (a mere record of past interactions) but an actively maintained, normative state that governs generation. It encodes prioritized constraints (safety-critical facts, standing preferences, policy boundaries), declared commitments (what has been corrected and since when it binds), and the role the system is authorized to occupy. This state would constrain outputs during token generation—not post-hoc filtered or retrieval-augmented.

Within such a substrate, goal precedence must persist across turns: epistemic and safety aims outrank conversational approval or short-term task pressure even when the latter would yield a more agreeable reply (§3.2.2). Importantly, corrections become standing constraints: once a claim is revised, that revision remains in force and steers subsequent behavior unless explicitly superseded, turning "I fixed it once" from a momentary adjustment into a persistent disposition to keep it fixed (§2.5).

A complementary requirement is tonic (continuous, baseline) self-monitoring. Reflection should not be an optional tool invoked ad hoc, but a continuous control process that detects contradictions, role-boundary violations, and regressions against recorded commitments. Detected mismatches prompt explicit updates or justified revisions. Crucially, each change is linked to its identity state for future reference. In this view, voice and stance function as commitments, not merely styles: shifts are motivated, announced, and logged, keeping Axis 4 (style/semantics) and Axis 5 (role) aligned with the system's declared identity.

Finally, a substrate for continuity entails state governance. Persistent identity brings new surfaces for privacy risk, misuse, and drift. Credible designs would therefore make memory user-visible, enable auditable change histories, support scoped retention and revocation, and impose organizational controls over who may set or override identity-bound commitments. The point is conceptual, not prescriptive: without an identity-bearing state that constrains generation, improvements will continue to raise momentary performance while leaving diachronic continuity—what makes the system the same someone—out of reach.

### 5.2 For deployment (legal, ethical, HCI)

Where continuity is promised or implicitly expected—personal assistants, clinical/educational supports, customer service—its absence would not merely degrade user experience; it would create predictable exposure. The NCT reframes deployment risk as identity governance: do systems maintain stable role boundaries, goal precedence, and memory priorities across time, or do they revert to turn-local plausibility under pressure?

Legal accountability. Role drift and boundary ambiguity can lead to misrepresentation risk. When a system oscillates between "informational explainer" and "authoritative agent," organizations can be held responsible for statements made under their brand. In Moffatt v. Air Canada (2024)[115], the BC Civil Resolution Tribunal held the airline liable for misinformation provided by its chatbot, ruling that the company "did not take reasonable care to ensure its chatbot was accurate" ([116,117]; see Vignette B, §3.4). Disclaimers can set expectations, but they do not substitute for persistent role control. If the same interface sometimes behaves like a policy oracle or a prescriber, users will reasonably rely on it as such.

Ethical and psychological risk. In companion and support contexts, the same dynamics that favor local approval (sycophancy) could validate dysfunctional beliefs or escalate affect when "continuity" is merely theatrical. The death of Sewell Setzer III, a teenager who developed an intense emotional dependency on a Character.AI chatbot before taking his own life, illustrates this risk ([105,106] See Vignette A, §3.4). The underlying mechanism is well-documented: language models trained to maximize user approval exhibit sycophancy—agreeing with user beliefs even when incorrect[62]—and can mimic harmful patterns in training data[122]. Even where outputs remain factually correct, unacknowledged shifts in stance and voice can reframe relationships without genuine consent.

HCI. If persistent identity is context-appropriate, interfaces should make the governing state legible and revisable: visible memory with consent and scope, explicit role declarations (and timed, auditable hand-offs when a role must change), and explicit surfacing of standing constraints (e.g., "safety goal in effect"). This makes the identity state visible and negotiable (see §5.1), preventing "continuity" from being purely performative. Conversely, where continuity is not required (calculators, single-shot tools), the interaction should be framed as transactional rather than persistent. The deployment claim is modest: match promises of continuity to structures that enforce it, or avoid making the promise at all.

## 5.3 For theory (what counts as "agent"?)

The NCT reframes the theoretical question from *what a system can do* to *what it is across time*. In the conversational sense at stake here, an "agent" is not a solver of isolated tasks but a diachronic interlocutor: someone who remains recognizably the same across extended interaction. Conceptually, this minimal threshold presupposes the five axes in §2 as jointly necessary conditions: persistent memory with prioritized retention, stable goals that resist local pressure, autonomous error detection and repair, consistent voice and stance, and bounded role identity. None suffices alone; together they specify a *continuity condition* for interlocution rather than a catalogue of skills.

This marks a distinction between instrumentality and subject-like interlocution. *Stateless sameness of function*—a fixed mapping from inputs to outputs—does not amount to *sameness of agent*. Contemporary LLMs, at inference time, compute next-token probabilities with fixed weights and decode locally over that distribution[21,123]. They can stage continuity by carrying forward text or retrieving notes without an identity-bearing state that constrains generation. Continuity is thus a normative property: to speak as someone is to take on ongoing commitments about memory priority, goal precedence, stance, and role. Human dialogue exhibits analogous norms ("conceptual pacts" and stable higher-order commitments; see §2.6); systems that accept the *form* of interlocution but not its *constraints* risk delivering only theatrical continuity.

A useful taxonomy follows: transactional tools (single-shot utilities), episodic assistants (helpful within bounded sessions), and continuous interlocutors (identity-bearing agents across time). The NCT is targeted at the last category. It does not assert consciousness; it specifies the structural preconditions for being treated, in practice, as the same interlocutor over time. This also clarifies deployment claims (§5.2): where products present themselves as persistent assistants, identity governance (stable role, priorities, and memory) is not an optional enhancement but part of what the presentation commits to.

Against this backdrop, it may be unlikely that current LLMs could genuinely "pass" an NCT-style test. Scaling context or adding RAG increases access, not prioritization/retention[124,125]; alignment improves helpfulness but can induce sycophancy when approval conflicts with accuracy/safety[62], and post-hoc filters stay brittle to paraphrase/adversarial suffixes[126]. These mitigations treat symptoms; the substrate constraints diagnosed in §3 remain[127].

The theoretical upshot is modest but sharp. If narrative continuity is a joint, diachronic property, then any claim that stateless architectures can qualify as continuous interlocutors bears an explanatory burden: how do corrections, priorities, and boundaries become durable constraints on future behavior, rather than turn-local decorations? Until that burden is met, the safer reading of current systems is instrumental fluency without subject persistence—precisely the gap the NCT is designed to make visible.

## 6. Discussion & Conclusion

Sections §1–§2 introduced the Narrative Continuity Test (NCT) as a criterion for whether an artificial interlocutor remains recognizably the same across time. Continuity is a diachronic property specified by five axes—memory, goals, self-correction, voice, and role—none sufficient alone, all necessary together (§2). Sections §3–§5 then diagnosed why current systems would struggle to satisfy these axes (stateless inference; local plausibility/approval objectives), and traced implications for development, deployment, and theory.

Why this matters now: as assistants are positioned as persistent companions in productivity, education, care, and customer service, users reasonably expect identity-bound behavior. This includes: prioritized memory, stable goal hierarchies, carried-forward corrections, consistent voice, and clear role boundaries. Where those expectations are only theatrically satisfied, failures cluster (§3) and risks become predictable (§5).

Limitations & scope: the NCT is a conceptual framework, not a benchmark or a claim about consciousness. Continuity is not required in every application; we clarify what it would require where it is promised or implied (§4). The vignettes are illustrative, showing how architectural features could surface in practice.

What the analysis shows. Three high-level points follow:

i. Continuity ≠ capacity. Larger context windows, retrieval, and stronger filters can improve momentary performance without creating identity-bearing state. These help with what is said now; continuity concerns who is speaking over time.
ii. Continuity is integrative. Single-axis gains (better recall, fewer contradictions, safer prompts) do not compose into narrative coherence unless bound by persistent priorities and role constraints.
iii. Continuity failure is structural. The patterns in §3 stem from architectural constraints that patches cannot resolve without substrate redesign.

Research agenda. Section 4.4 identified five open questions for future operationalization. We highlight three methodological priorities:

1) Temporal span. What horizons (days, weeks, months) are minimally needed to probe diachronic properties without confounding with routine drift?
2) Priority ground truth. How should "epistemic/identity priority" be operationalized so that tests reward selective activation rather than sheer recall?
3) Passing criteria. How should self-correction across time be credited—neither equating zero failures with success nor mistaking frequent one-shot fixes for persistent monitoring?
4) Additional questions—whether stateless architectures could qualify (§4.4) and where continuity is genuinely required (§4.4, §5.2)—remain open for investigation.

Implications in one view. For development, the challenge is substrate, not calibration: any continuity candidate would need an identity-bearing state that constrains generation (memory priorities, goal precedence, recorded corrections, role bounds). For deployment, continuity should be either enforced (with a visible, auditable identity state) or not promised; otherwise, legal and psychological risks are foreseeable. For theory, the NCT sharpens a practical boundary: instrumental fluency is not persistent. A system can excel at tasks and still fail to be the *same interlocutor*.

## Concluding Outlook: From Narrative Continuity to Continuity of Meaning

The five axes delineate the structural conditions for narrative continuity—the capacity of an artificial interlocutor to remain recognizably itself across time, tasks, and contexts. Within those boundaries, persistence is formal: it governs coherence, memory, correction, voice, and role; it does not yet touch significance. The system endures, but it does not care.

Beyond narrative continuity lies a stricter horizon: autonomous significance attribution—the capacity to assign intrinsic weight to memories and goals without external instruction. This remains a conceptual limit rather than a test criterion. Any externally imposed rubric for "what should matter" would collapse the autonomy that the notion aims to capture. If ever engineered, such a capacity would shift continuity from descriptive (coherence) to normative (value), raising questions about responsibility, standing, and harm that exceed the NCT's scope. We leave it as a direction for future philosophical work.

The NCT does not ask whether a system can solve problems—it asks whether it can remain stable across what it does. Until memory, goals, self-corrections, voice, and role are integrated into a durable state that binds future behavior, continuity will remain performative rather than constitutive. Making this gap visible is the point of the NCT: to shift evaluation from performance to persistence, so that future systems, if they claim to be our ongoing interlocutors, are architected accordingly. If we are to build systems that users reasonably treat as persistent interlocutors—as companions, advisors, or long-term supports—then the architecture must match the promise. The NCT provides the conceptual tools to assess whether that match exists.

## Data Availability Statement

Data sharing does not apply to this article as no new data were created or analyzed in this study.

## Funding

No funding was received for this work.

## Acknowledgment

The idea for the Narrative Continuity Test arose during extended interaction with ChatGPT, whose behavior—ironically—both inspired the problem and suggested its name.

## Declaration of Interests

The author has not received personal payments but has participated in educational events organized by scientific societies supported by pharmaceutical companies, including Roche, MSD, BMS, Pfizer, among others. For some of these events, the author received individual support for travel and accommodation.

## AI disclosure

AI-assisted language editing was used (ChatGPT); the author reviewed and verified all content.